\documentclass[fleqn,10pt]{wlscirep}
\usepackage[utf8]{inputenc}
\usepackage[T1]{fontenc}

\usepackage{sidecap}

\title{Imitating the winner leads to discrimination}

\author[*]{Gorm Gruner Jensen}
\author[ ]{Stefan Bornholdt}
\affil[ ]{Institute for Theoretical Physics, University of Bremen, Bremen, 28359, Germany}

\affil[*]{ggjensen@itp.uni-bremen.de}

\begin{abstract}
The occurrence of discrimination is an important problem in the social and economical sciences.
Much of the discrimination observed in empirical studies can be explained by the theory of in-group favoritism, 
which states that people tend to act more positively towards peers whose appearances are more similar to their own.
Some studies, however, find hierarchical structures in inter-group relations, where members of low-status groups 
also favor the high-status group members.
These observations cannot be understood in the light of in-group favoritism.
Here we present an agent based model in which evolutionary dynamics can result in a hierarchical discrimination 
between two groups characterized by a meaningless, but observable binary label.
We find that discriminating strategies end up dominating the system when the selection pressure is high,
i.e.\ when agents have a much higher probability of imitating their neighbor with the highest payoff.
These findings suggest that the puzzling persistence of hierarchical discrimination 
may result from the evolutionary dynamics of the social system itself, namely the social imitation dynamics. 
It also predicts that discrimination will occur more often in highly competitive societies.
\end{abstract}
\begin{document}

\flushbottom
\maketitle
\thispagestyle{empty}

\section*{Introduction}
\label{sec:Introduction}

Structural discrimination is a problem across a wide range of societies. 
Often, there is no apparent connection between the defining characters of a group 
and an obvious rational reason to discriminate against its members.
One of the most commonly observed discriminating behaviors \cite{Hewstone2002, Efferson2008}, 
in-group favoritism, is characterized by a tendency to show more cooperation, preference, 
or altruistic behavior towards people whose appearance is close to ones own (the in-group) 
than to those who appear different (the out-group).
However, a growing body of research seems to indicate that members of some minority groups, 
particularly those of low social status, favor their in-group much less than members of 
high status groups, or even favor members of their high-status out-group (in-group devaluation) \cite{Ridgeway1998,Jost2000,Jost2002,Rudman2002,Jost2004,becker2010women,March2014,Kaiser2015,Proestakis2016}. 
In-group favoritism clearly cannot explain why such hierarchies should be accepted by those 
they disadvantage. This has lead to the development of system justification theory, which states 
that people have a intrinsic tendency to legitimize and preserve systemic forms of inequality 
\cite{Jost1994}.
System justification theory provides an efficient explanation for the observed out-group favoritism 
from low-status group to high-status groups by introducing a fairly strong assumption about human psychology.
This raises the question about whether this kind of behavior could be explained from simpler, 
or more fundamental, principles.

Some investigations following the classical economic principle of selfish rationality have 
revealed an important connection between discrimination and incomplete information.
A rational agent who has to choose between individual options belonging to different 
groups\footnote{this could be an employer choosing between different employees, or a 
	customer choosing between different suppliers} may try to compensate incomplete information 
about the quality of each choice by factoring in a prior knowledge about the quality 
distributions within each group.
If the quality-distributions of the two groups are very different this knowledge is weighted heavily, 
it could lead the agent to choose an option with a weaker individual performance estimate, 
but belonging to an -- on average -- stronger group, instead of someone with a stronger 
individual performance estimate, but belonging to a weaker group \cite{Phelps1972}.
This statistical theory of discrimination has been used as a fundamental building block 
for designing dynamical models in which the collective reputation of group will be trapped 
in one of a number of possible Nash-equilibria stabilized by positive-feedback between 
the expected and the optimal behavior \cite{Tirole1996, Levin2009}.
These types of models suggest that persistent discrimination can be explained without 
assuming any differences of intrinsic properties of the members of different groups. 
All the existing models of this kind, however, describe asymmetric interactions, 
where the `discriminating' agent belongs to a completely different category than 
the `discriminated' agents. Thus, none of them directly describes in-group devaluation.
Also, there may not be a simple way of introducing it in this framework, because 
a rational agent would be expected to recognize that in-group devaluation would negatively 
affect its own future possibilities, and avoid acting against its own self-interest.
One way to circumvent this problem could be to loosen the assumption of rationality. 
This could be further motivated if the discriminating behavior often occurs unconsciously, 
or indirectly via complex social interactions which are not fully understood by the agents.

Evolutionary game theory is one framework which has been enormously effective 
in explaining social phenomena such as altruism and cooperation \cite{Nowak2006}.
In contrast to the classical economical theories it approaches the behavioral question 
starting with almost completely irrational agents who do not have any direct knowledge 
about the consequences of their actions. Rather than making assumptions about the 
actual decision process, it works by assuming that behaviors are fundamentally random, 
but successful strategies are promoted by replicating at a higher rate.
This replication is commonly interpreted either as biological reproduction of genes or,
in the context of cultural evolution, as mimicking of ideas or behavioral motives.
A number of studies have investigated how the introduction of more or less arbitrary tags can be used to promote the evolution of cooperation 
\cite{Riolo2001, Traulsen2007, Antal2009, Fu2012, Hadzibeganovic2014, Zhang2015}.
More recently a study has investigated the effect of tags when considered in combination with population structure --- another mechanism well known to promote cooperation \cite{Garcia2014}.
There it was shown that the introduction of tags could also reduce the overall level of cooperation, by allowing discriminating strategies to invade a population of unconditional cooperators. 
It has been pointed out by Fu et al. \cite{Fu2012} that this mechanism 
can be used to explain the evolution of in-group favoritism within dynamic groups 
where memberships may change quickly, e.g. political movements.
The existing models of tag-based cooperation do, however, not reproduce predict the observed phenomenon of sustained hierarchical discrimination, i.e. a mixed population in which members of one group are treated preferentially, both by their peers, as well as by members of the less fortunate groups.

Here we suggest a new model for investigating hierarchical discrimination in 
the framework of evolutionary game theory.
Starting from an established model describing the evolution of cooperation in a 
prisoner's dilemma type game on graphs \cite{Ohtsuki2006}, we partition the population 
into two groups by randomly assigning an observable, but completely meaningless, 
binary label to each agent (blue or green).
As the labels are observable, this expands the set of possible strategies from two 
(\textit{cooperate} and \textit{defect}), to four (1: \textit{cooperate with all}, 
2: \textit{cooperate only with blue}, 3: \textit{cooperate only with green}, 
4: \textit{defect all}).
We call strategies discriminating when they imply different behavior towards 
agents carrying different labels.

In contrast to models of tag-based cooperation, the labels described in this model never change.
Hence the only dynamical variable in the model are the strategies, which agents change mainly by imitating their neighbours.
As strategies spread from agent to agent, the model has a tendency to let one 
strategy dominate all (or at least large regions) of the population.
Thus our main task is to investigate which conditions (choices of parameters) promotes the dominance of discriminating strategies.

\section*{Model}

\label{sec:model}
\begin{SCfigure}[][b!]
	\includegraphics[width=.45\linewidth]{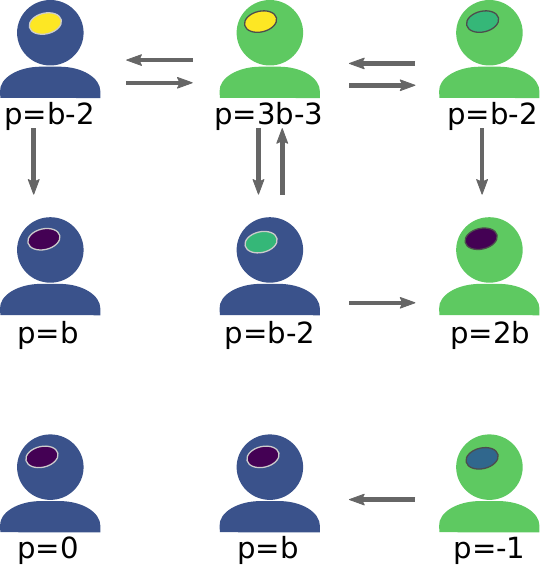}
	\caption{
		\label{fig:modelDescription}
		Model description: 
		Agents are located on a square lattice. 
		Each agent has a label (blue or green) which can be observed by their neighbors, 
		and a strategy (yellow: "cooperate with all", green: "cooperate with green", 
		blue: "cooperate with blue", purple: "defect all") which determines their 
		interaction with their neighbors. From this configuration, an agent's payoff $p$ 
		is calculated by subtracting $1$ for each neighbor it cooperates with, 
		and adding $b$ for each neighbor cooperating with it. Here the parameter $b$ 
		represents the benefit of cooperation. Each time-step one randomly chosen agent 
		changes its strategy by copying that of one of its neighbors.
		This neighbor is chosen with a probability proportional to $f=\exp(wp)$, 
		where the parameter $w$ represents the selection pressure.
		When the selection pressure is small ($w \rightarrow 0$) neighbors are 
		chosen with almost equal probability independent of their payoff.
		When the selection pressure is high ($w \rightarrow \infty$) the 
		neighbor with the highest payoff will almost certainly be chosen.
	}
\end{SCfigure}

Let us define a game of agents distributed on a graph. Each agent has a static and binary 
`label', either green or blue, decided at the beginning of the game.
This label serves as the only observable difference between agents. 
Agents interact with their nearest neighbors in a prisoner's dilemma type of game where 
they can either cooperate or defect. When cooperating, an agent donates one unit of 
value to the neighbor. To simulate the benefit of cooperation, the donation is scaled 
by a constant factor $b>1$, such that the neighbor receives $b$ times the unit value.
When the an agent is defecting, no value is transferred.
Since agents cannot distinguish between neighbors who have the same label they must 
act the same way towards all of them. Thus the model has four possible strategies: 
``defect all'', ``cooperate with all'', ``cooperate only with green'', and 
``cooperate only with blue''.

Given a configuration of labels and strategies, a payoff, $p$, can be calculated for each agent. 
An agent's payoff is the sum of all donations the agent is receiving from its neighbors, 
minus all the donations it is giving away. Notice that the payoff of an agent $i$ 
does not accumulate over time, and can always be calculated from the current state 
by: $$ p_i = \sum_{j\in\mathcal{N}_i} b\cdot S_j(\lambda_i) - S_i(\lambda_j), $$, 
where $\mathcal{N}_i$ is the set of neighbors of agent $i$, 
$\lambda_i\in\{\text{blue, green}\}$
 is the label of agent $i$,
and $S_i(\lambda_j) = 1$ if the strategy of agent $i$ is to cooperate with the label 
worn by agent $j$, and $S_i(\lambda_j) = 0$ if it is to defect.

The evolutionary dynamics of the spreading of the strategies is as follows.
Every turn a random agent is chosen (from a uniform distribution over all agents).
With a small probability $\mu$, this agent will mutate, i.e.\ choose a new random 
strategy (from a uniform distribution over all strategies).Else, with probability 
$1-\mu$, the chosen agent will imitate the behavior of one of its neighbors.
In that case, a neighbor is chosen with a probability proportional to its fitness $f$ 
which is directly related to its payoff by: $$ f_i = e^{w\cdot p_i} $$
where $f_i$ and $p_i$ are the fitness and payoff of agent $i$, and $w$ is 
a model parameter controlling the selection pressure. When the selection pressure 
approaches zero, all neighbors are chosen with almost equal probability.
When the selection pressure is high, the neighbor with the highest payoff 
will almost certainly be chosen.

Our model is characterized by three parameters: The cooperation benefit $b$, 
the selection pressure $w$, and the mutation rate $\mu$.

The random mutations serve two functions: Adding noise and preventing strategies from going extinct. 
We are interested in the evolutionary dynamics of the imitation dynamics, and not in the noise.
Thus we keep the mutation-rate small and constant ($\mu = 0.001$) throughout this paper.

Besides the three parameters mentioned above, we have two `hidden' parameters, namely the underlying graph topology and the distribution of labels. 
In this paper we have chosen to focus on a 2-dimensional square lattice,
because it allow for intuitive visualization of the model-state.
Our lattice has periodic boundary conditions in one direction, to form a cylinder.
The labels are then randomly distributed such that the probability that an agent is given a green label increases linearly from zero in one end of the cylinder (`top'), to one in the other end (`bottom').
This allows us to effectively investigate the effect which emerge due to a mesoscale imbalance in the label density.

We refer interested readers to the supplement for results with uniform label-distribution on a 2D lattice \ref{fig:uniform-scan}, Erd\H{o}s-R\'{e}nyi random graphs \ref{fig:random-graphs}, and a 1D ring-topology \ref{fig:1d-scan}. 
The parameter scans show qualitatively similar results, indicating that the behaviour doesn't depend critically on the topology.

The model described here is an extension of a model introduced by Ohtsuki et al.\ 
in an effort to study how spacial structures may promote evolution of cooperation 
\cite{Ohtsuki2006}.
The most important change is the introduction of the observable labels.
However, we also differ by defining the fitness using using the exponential 
function $ f_i = e^{w\cdot p_i} $, whereas they choose $f= 1-w+w\cdot p$.
In the limit of low selection pressure ($w\rightarrow0$), these two definitions converge. 
The choice of the exponential function, however, secures that all fitnesses 
will be positive no matter how high the selection pressure. This allows us to 
investigate all the way to the deterministic limit, where agents always imitate

\section*{Results}
\label{sec:results}

\begin{figure*}[t]
	\includegraphics[width=\linewidth]{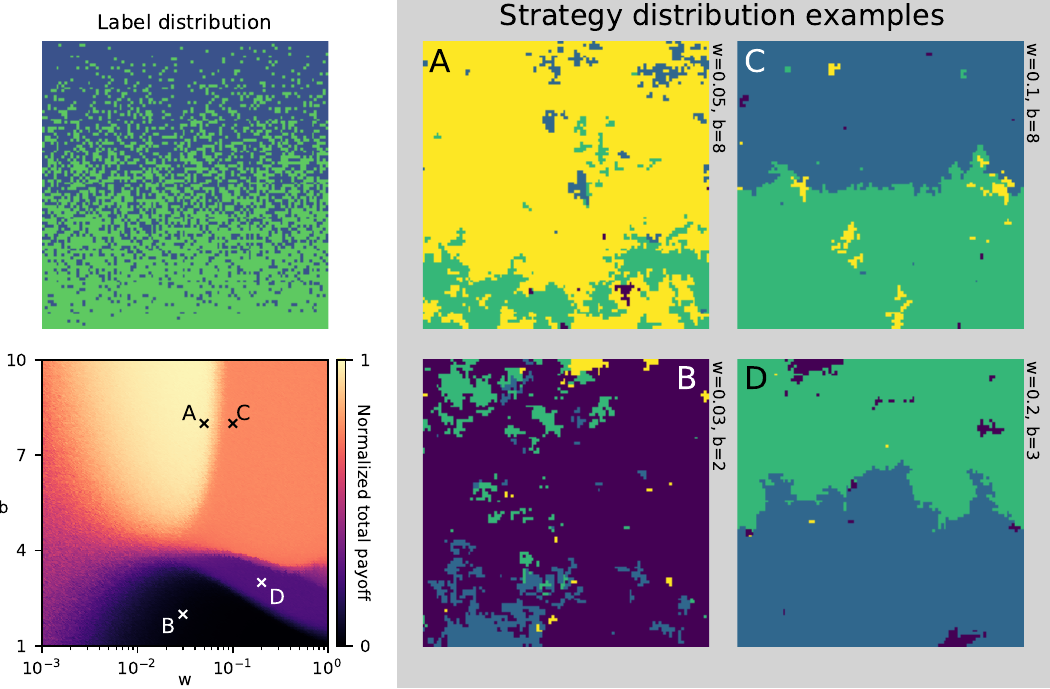}
	\caption{ \label{fig:phaseDiagram}
		Phase-diagram. Constant mutation rate, $\mu = 0.001$, and grid-size, $100\times100$ agents.
		\textbf{Top-left:} The label distribution used in each of the examples to the right. 
		In the parameter scan, a new label distribution is generated at every point, 
		to ensure that the results are not unexpectedly caused by random local structures.
		\textbf{Bottom-left:} Parameter scan over cooperation benefit $b$, and selection pressure $w$.
		The color indicates the mean payoff normalized, 
		averaged over 20 samples uniformly distributed over a period of $10^8$ time steps.
		At each data point the system is initialized with all defectors and run for a transient period 
		of $2\times10^{7}$ time steps before the mean payoff is measured.
		\textbf{Right:} Four snapshots of strategy-distributions at parameters corresponding to 
		those marked in the parameter scan.
	}
\end{figure*}

Figure~\ref{fig:phaseDiagram} captures the long-term behavior of the model, 
as it settles into different stationary states dependent on parameters.

In the bottom left panel the normalized mean payoff is plotted (in color) as 
a function of the selectionn pressure $w$ and the cooperation benefit $b$.
The normalized mean payoff is identical to the fraction of donations in the 
system out of all possible ones. Due to the stochastic nature of the model, 
the system is never locked in an absorbing fixed point. The normalized mean payoff, 
however, tends to stabilize after some transient period, as the system finds 
a stationary distribution of strategies. To capture this long term behavior, 
we initialized the system with all defectors and run the model for a 
transient period of $2\times10^{7}$ time steps before the mean payoff is measured.
We then measure the averaged value over 20 samples evenly spaced over a period of $10^8$ time steps.
The phase diagram clearly shows four distinct regions characterized by different levels of cooperation.

In the right panel, four examples of stationary state strategy distributions are shown, one for each phase.
These snapshots show the following characteristics, which we have observed consistently throughout all our simulations:

\textbf{A:} The system is dominated by full cooperation resulting in close to maximal mean payoff.
As expected, this optimal state requires that the cooperation benefit is greater than 
4 (average connectivity), as predicted by Ohtsuki et al.'s simple rule for  
the evolution of cooperation on graphs and social networks \cite{Ohtsuki2006}.
Surprisingly, we find that it is also necessary to have a relatively low 
selection pressure to obtain this state of almost full cooperation.

\textbf{B:} When the cooperation benefit is low, the dominant strategy is complete defection, 
resulting in approximately zero payoff. Surprisingly we find that when the selection pressure 
is high ($w \gtrsim 0.03$), full defection is outperformed by strategies with non-zero 
cooperation even when the cooperation benefit is less than four (the average connectivity).

\textbf{C:} In general, when the selection pressure is high, we observe that the system is dominated by the asymmetric, or discriminating, strategies. 
The result is the formation of regional hierachies, where all agents cooperate with neighbors carrying one of the labels, independent of their own label.
Thus the `upperclass' benefit from the cooperation from all of their neighbours while only donating to their in-group.
The exploited agents, on the other hand, display in-group devaluation by only cooperating with members of their out-group.

These hierachical regions have an interesting connection with the local label densitiy.
With our choice of label distribution, the lattice of agents is split into two qualitatively different regions.
In `the top' of the cylinder, the majority of agents have the blue label, while in `the bottom' the green label is much more common. 
When the cooperation benefit is high, $b\gtrsim 4$ (average connectivity), we find a threshold for the selection pressure, above which the dominating strategy within each of the two domains (`top' and `bottom') is to cooperate with the local majority label, while defecting against those in the minority. 
As a result, the normalized mean payoff is approximately equal to the population fraction of the majority in each domain, i.e.\ $3/4$ given this specific label-distribution.


\textbf{D:} For high selection pressures and intermediate selection benefits $b$ slightly below 4 (the average connectivity), 
we observe that domains of the asymmetric cooperation strategies have switched around as  
compared to the case with larger cooperation benefit $b > 4$ (the average connectivity).
Cooperating with agents carrying one label, and defecting those carrying another, still out-competes both of the symmetric strategies, but within each of the two domains (`top' and `bottom') 
the agents in the majority are defected, and only those in the minority group receive donations.
Consequently, the normalized mean payoff is equal to the fraction of minority labels 
within each domain which is $1/4$ with the given label-distribution. 

Closer examination of the parameter scan in figure~\ref{fig:phaseDiagram} reveals 
that the transition between the different phases are qualitatively different.
For low selection pressure (left side of the figure), there is a smooth transition 
between `low mean payoff' when the cooperation benefit is low, to `high mean payoff' 
when the cooperation benefit is high. For intermediate cooperation benefits, we find 
very noisy stationary distributions in which patches of all possible strategies coexist. 
These high-entropy distributions are easy to understand. 
Because of the very low selection pressure the boundaries between strategy patches 
perform a random walk almost without a drift. When $w=0$ dynamics are equivalent 
to a classical voter-model \cite{cox1986diffusive,dornic2001critical} with four opinions.

The sharp phase transitions observed at higher selection pressures ($w\gtrsim 0.03$) 
are more intriguing because 
1: The sharp transitions indicate that relatively small changes in the environment 
can have dramatic effects on the social structure, and
2: The sharp transitions separate the regions dominated by discriminating strategies.

The most intriguing finding in this new model, is the symmetry breaking phase transition 
in which full cooperation is ousted by a strategy of selectively cooperating with 
agents carrying one of the labels but not with those carrying the other label. Surprisingly, 
we find that, as long as the cooperation benefit is high (somewhat higher than the average 
connectivity), transition-point is determined almost solely by the selection pressure.
In fact it seems that for selection pressures above a certain threshold, full cooperation 
cannot be restored no matter how much the cooperation benefit is increased.
To get a better understanding of this, we have performed a detailed study of a simplified scenario.

In this example, all agents are arranged on a one-dimensional line rather than a square lattice.
Each agent has only two neighbors, and all agents except one have the blue label. The single 
agent with the green label is placed in the middle of the line, and the mutation-rate is set to zero. 
Instead we fix the strategy of the outermost agent in one end of the line to always being ``full cooperation'',
and the strategy of the outermost agent in the other end to ``cooperating with blue, but defecting green''.
Any strategy squeezed in between these two will eventually disappear (by random fluctuation), 
and the complete state of the system can be described by one number indicating the location 
of the boundary between one strategy and the other.
Notice that outside the neighborhood of the one green agent, these two strategies lead to 
identical behavior, and thus the boundary between them will simply make an unbiased random walk.
In the within 3 steps of the green agent, however, the boundary will move left or right with 
a probability that must be calculated explicitly for each position.
After having determined the individual stepping probabilities we calculate the probability of finding it on the symmetric side of the green agent (the asymmetric strategy dominates) and 
the probability of finding it on the asymmetric side (the symmetric strategy dominates), assuming a stationary distribution:
$$	 \frac{\mathit{P}(\text{discrimination dominates})}{\mathit{P}(\text{full cooperation dominates})}= 
\frac{e^{2w}}{\tanh{(wb)}+1}
$$
When the fraction is greater than one, we expect a strategy of asymmetric cooperation to be able to outperform the full cooperation in a mixed population. Comparison with the 
phase diagram for the one-dimensional system show a close, but non-perfect match 
(see supplementary figure~\ref{fig:1d-scan}). In the limit of very high cooperation-benefit, $b\gg 1$, the condition reduces to $w>\log(2)/2$. This result confirms the visual impression from the parameter-scan, that for high enough selection pressures, discriminating strategies dominate no matter how high the cooperate benefit is.

More details about the calculations can be found in the supplementary material.

\section*{Discussion}

In this paper, we have introduced and studied a minimal model in which evolutionary dynamics can promote discriminating behavior between arbitrary labels resulting in a reduction of the overall level of cooperation.
Our model describes the cultural evolution of competing behaviors/strategies, 
of which some are discriminating, in the sense that they imply different behaviors 
towards agents carrying different labels.
We have observed that the discriminating strategies end up dominating the system 
when the selection pressure is high -- that is, when agents have a much higher 
tendency to imitate the behavior of their most successful neighbor -- resulting 
in a hierarchical society where agents carrying one label end up with a higher 
payoff than those carrying the other.
This hierarchy emerges spontaneously even though the model treats the labels 
symmetrically, and there is no extrinsic preference for choosing 
one strategy over any other. 
In fact, agents choose their strategy completely independently of their label.
As a consequence the low status agents (those with the lowest payoff) end up exhibiting out-group favoritism, by cooperating only with those neighbors having a different label than their own.
This type of behaviour is in qualitative agreement with experimental data suggesting that members of certain low-status groups tend to express significantly lower in-group favoritism, or even favoring high-status group members \cite{Ridgeway1998,Jost2000,Jost2002,Rudman2002,Jost2004,becker2010women,March2014,Kaiser2015,Proestakis2016}.
These observations are usually explained with system justification theory \cite{Jost1994}.
System justification theory states that humans have an intrinsic drive for justifying -- and thereby validating -- established structures around them.
In contrast to the well-established psychological theory of system justification, our model is a minimalistic toy-model.
We have ignored many important aspects of human nature, as well as making unrealistic assumptions, in an attempt to make the model as simple as possible, while still qualitatively capturing the observed phenomenon.
Our model is intended as an abstract investigation of the idea, that mechanisms for hierarchical discrimination could be initiated or reinforced via group-dynamical mechanisms outside of individual preferences.
We do not, however, understand this as a replacement for the established theory, but rather as a complimentary mechanism.

It is worth noting, that the behavior observed in the model described in this paper, differs in two fundamental aspects from that described in the large literature of tag-based cooperation.
Firstly, we find a strong tendency for our system to settle into a  stationary macro-distribution of strategies. This is in sharp contrast to the oscillatory (or wave-like) behaviour of previous models \cite{Traulsen2007, Antal2009, Fu2012, Hadzibeganovic2014, Garcia2014, Zhang2015} (\cite{Riolo2001}). In these models, the system is almost always susceptible to invasion from new strategies. When the population is dominated by complete defection, it is vulnerable to the random emergence of an agent with a currently non-existing tag and an in-group favoring strategy. Later, when that agent-type has taken over most of the population, it becomes susceptible to free-riders who have the same tag, but do not cooperate with their peers.

Secondly, when discriminating strategies become dominant in models of tag-based cooperation, they do so by the fortune of a strong correlation between a given tag and an in-group favoring strategy. 
When we observe discriminating strategies dominating in our model, it tends to form large (mesoscale) regions each dominated by a single strategy --- either ``cooperate only with green'' or ``cooperate only with blue''. 
Within each of these regions, the population is generally a mix of agents with different labels. 
Consequently a non-negligible fraction of the population is actually expressing out-group favoritism.

The key to understanding these differences lies in the implementation of the evolutionary algorithm.
In models of tag-based cooperation, tags are passed on to future generation together with the strategy of the parent, unless a rare mutation occurs.
A successful combination of tag and strategy will therefore quickly invade the system, resulting in a population where a large majority of agents have the same tag.
In the model presented in this paper, agents never change their label, and agents with one label can adopt the strategy of neighbours with the other label.
Thus a strategy is able to produce the richest individuals by letting the agents carrying one label exploit those carrying the other.
This also helps explaining the observed connection between a high selection pressure, $w$, and the dominance of discriminating strategies. 
The selection pressure can be interpreted as the agent's `eagerness' to copy their richest neighbour.
When it is strong enough the contagion benefit of strategies that produce richer agents outperforms the disadvantage of also producing poor ones.
If we keep this analysis in mind we can also argue, that a strategy with explicit in-group favoritism (``cooperate only with neighbours carrying the same label as myself'') would not have an evolutionary advantage under these dynamics.
With such a strategy no agents would have the advantage of receiving donations from their neighbours without giving back.

The models of discrimination which have followed a classical 
economical approach by assuming rational agents 
\cite{Tirole1996, Levin2009} also cannot capture the cause of 
discrimination that we have observed in our model.
This is because it is crucial to the development of hierarchical 
discrimination in our model that agents do not act intelligently 
in the direction of self-interest.
Even in economical models where agents have imperfect information 
about each other, they are typically assumed to at least have full 
information about the fundamental rules of the game they are playing.
It can be argued that evolutionary game theory makes the general 
assumption that agents have no understanding about cause and effect 
in their social interactions, and thus must rely on observed 
correlations between other agents' behavior and their profit, 
in order to try to optimize their game.
Such an assumption seems more likely in the context of unconscious 
behaviors, or in complex social interactions where the causal 
connection between actions and delayed rewards is in fact indirect.

Many interesting questions remain in the context of the non-group-based 
mechanism of discrimination in social systems that we studied in 
a minimal model in this paper.
In particular, we have enforced a spatial structure on the model-population which is one out of a number of well known ways to promote cooperation in prisoner's dilemma type games \cite{Ohtsuki2006, Nowak2006}.
At this point it remains an open question whether a cooperation-reducing, 
hierarchical discrimination will also emerge via spontaneous symmetry-breaking 
when introducing meaningless tags into models where cooperation is 
achieved by different means.

\section*{Acknowledgements}
This is a pre-print of an article
published in Scientific Reports. The final authenticated version
is available online at: https://doi.org/10.1038/s41598-019-40583-w
%

\bibliography{biblo}

\begin{thebibliography}{10}
\expandafter\ifx\csname url\endcsname\relax
  \def\url#1{\texttt{#1}}\fi
\expandafter\ifx\csname urlprefix\endcsname\relax\def\urlprefix{URL }\fi
\providecommand{\bibinfo}[2]{#2}
\providecommand{\eprint}[2][]{\url{#2}}

\bibitem{Hewstone2002}
\bibinfo{author}{Hewstone, M.}, \bibinfo{author}{Rubin, M.} \&
  \bibinfo{author}{Willis, H.}
\newblock \bibinfo{title}{Intergroup bias}.
\newblock \emph{\bibinfo{journal}{Annual Review of Psychology}}
  \textbf{\bibinfo{volume}{53}}, \bibinfo{pages}{575--604}
  (\bibinfo{year}{2002}).
\newblock
  \urlprefix\url{https://doi.org/10.1146/annurev.psych.53.100901.135109}.

\bibitem{Efferson2008}
\bibinfo{author}{Efferson, C.}, \bibinfo{author}{Lalive, R.} \&
  \bibinfo{author}{Fehr, E.}
\newblock \bibinfo{title}{The coevolution of cultural groups and ingroup
  favoritism}.
\newblock \emph{\bibinfo{journal}{Science}} \textbf{\bibinfo{volume}{321}},
  \bibinfo{pages}{1844--1849} (\bibinfo{year}{2008}).
\newblock \urlprefix\url{https://doi.org/10.1126/science.1155805}.

\bibitem{Ridgeway1998}
\bibinfo{author}{Ridgeway, C.~L.}, \bibinfo{author}{Boyle, E.~H.},
  \bibinfo{author}{Kuipers, K.~J.} \& \bibinfo{author}{Robinson, D.~T.}
\newblock \bibinfo{title}{How do status beliefs develop? the role of resources
  and interactional experience}.
\newblock \emph{\bibinfo{journal}{American Sociological Review}}
  \bibinfo{pages}{331--350} (\bibinfo{year}{1998}).

\bibitem{Jost2000}
\bibinfo{author}{Jost, J.~T.} \& \bibinfo{author}{Burgess, D.}
\newblock \bibinfo{title}{Attitudinal ambivalence and the conflict between
  group and system justification motives in low status groups}.
\newblock \emph{\bibinfo{journal}{Personality and Social Psychology Bulletin}}
  \textbf{\bibinfo{volume}{26}}, \bibinfo{pages}{293--305}
  (\bibinfo{year}{2000}).
\newblock \urlprefix\url{https://doi.org/10.1177/0146167200265003}.

\bibitem{Jost2002}
\bibinfo{author}{Jost, J.~T.}, \bibinfo{author}{Pelham, B.~W.} \&
  \bibinfo{author}{Carvallo, M.~R.}
\newblock \bibinfo{title}{Non-conscious forms of system justification: Implicit
  and behavioral preferences for higher status groups}.
\newblock \emph{\bibinfo{journal}{Journal of Experimental Social Psychology}}
  \textbf{\bibinfo{volume}{38}}, \bibinfo{pages}{586--602}
  (\bibinfo{year}{2002}).
\newblock \urlprefix\url{https://doi.org/10.1016/s0022-1031(02)00505-x}.

\bibitem{Rudman2002}
\bibinfo{author}{Rudman, L.~A.}, \bibinfo{author}{Feinberg, J.} \&
  \bibinfo{author}{Fairchild, K.}
\newblock \bibinfo{title}{Minority members' implicit attitudes: Automatic
  ingroup bias as a function of group status}.
\newblock \emph{\bibinfo{journal}{Social Cognition}}
  \textbf{\bibinfo{volume}{20}}, \bibinfo{pages}{294--320}
  (\bibinfo{year}{2002}).
\newblock \urlprefix\url{https://doi.org/10.1521/soco.20.4.294.19908}.

\bibitem{Jost2004}
\bibinfo{author}{Jost, J.~T.}, \bibinfo{author}{Banaji, M.~R.} \&
  \bibinfo{author}{Nosek, B.~A.}
\newblock \bibinfo{title}{A decade of system justification theory: Accumulated
  evidence of conscious and unconscious bolstering of the status quo}.
\newblock \emph{\bibinfo{journal}{Political Psychology}}
  \textbf{\bibinfo{volume}{25}}, \bibinfo{pages}{881--919}
  (\bibinfo{year}{2004}).
\newblock \urlprefix\url{https://doi.org/10.1111/j.1467-9221.2004.00402.x}.

\bibitem{becker2010women}
\bibinfo{author}{Becker, J.~C.}
\newblock \bibinfo{title}{Why do women endorse hostile and benevolent sexism?
  the role of salient female subtypes and internalization of sexist contents}.
\newblock \emph{\bibinfo{journal}{Sex Roles}} \textbf{\bibinfo{volume}{62}},
  \bibinfo{pages}{453--467} (\bibinfo{year}{2010}).

\bibitem{March2014}
\bibinfo{author}{March, D.~S.} \& \bibinfo{author}{Graham, R.}
\newblock \bibinfo{title}{Exploring implicit ingroup and outgroup bias toward
  hispanics}.
\newblock \emph{\bibinfo{journal}{Group Processes {\&} Intergroup Relations}}
  \textbf{\bibinfo{volume}{18}}, \bibinfo{pages}{89--103}
  (\bibinfo{year}{2014}).
\newblock \urlprefix\url{https://doi.org/10.1177/1368430214542256}.

\bibitem{Kaiser2015}
\bibinfo{author}{Kaiser, C.~R.} \& \bibinfo{author}{Spalding, K.~E.}
\newblock \bibinfo{title}{Do women who succeed in male-dominated domains help
  other women? the moderating role of gender identification}.
\newblock \emph{\bibinfo{journal}{European Journal of Social Psychology}}
  \textbf{\bibinfo{volume}{45}}, \bibinfo{pages}{599--608}
  (\bibinfo{year}{2015}).
\newblock \urlprefix\url{https://doi.org/10.1002/ejsp.2113}.

\bibitem{Proestakis2016}
\bibinfo{author}{Proestakis, A.} \& \bibinfo{author}{Bra{\~{n}}as-Garza, P.}
\newblock \bibinfo{title}{Self-identified obese people request less money: A
  field experiment}.
\newblock \emph{\bibinfo{journal}{Frontiers in Psychology}}
  \textbf{\bibinfo{volume}{7}} (\bibinfo{year}{2016}).
\newblock \urlprefix\url{https://doi.org/10.3389/fpsyg.2016.01454}.

\bibitem{Jost1994}
\bibinfo{author}{Jost, J.~T.} \& \bibinfo{author}{Banaji, M.~R.}
\newblock \bibinfo{title}{The role of stereotyping in system-justification and
  the production of false consciousness}.
\newblock \emph{\bibinfo{journal}{British Journal of Social Psychology}}
  \textbf{\bibinfo{volume}{33}}, \bibinfo{pages}{1--27} (\bibinfo{year}{1994}).
\newblock \urlprefix\url{https://doi.org/10.1111/j.2044-8309.1994.tb01008.x}.

\bibitem{Phelps1972}
\bibinfo{author}{Phelps, E.~S.}
\newblock \bibinfo{title}{The statistical theory of racism and sexism}.
\newblock \emph{\bibinfo{journal}{The american economic review}}
  \textbf{\bibinfo{volume}{62}}, \bibinfo{pages}{659--661}
  (\bibinfo{year}{1972}).

\bibitem{Tirole1996}
\bibinfo{author}{Tirole, J.}
\newblock \bibinfo{title}{A theory of collective reputations (with applications
  to the persistence of corruption and to firm quality)}.
\newblock \emph{\bibinfo{journal}{The Review of Economic Studies}}
  \textbf{\bibinfo{volume}{63}}, \bibinfo{pages}{1} (\bibinfo{year}{1996}).
\newblock \urlprefix\url{https://doi.org/10.2307/2298112}.

\bibitem{Levin2009}
\bibinfo{author}{Levin, J.}
\newblock \bibinfo{title}{The dynamics of collective reputation}.
\newblock \emph{\bibinfo{journal}{The B.E. Journal of Theoretical Economics}}
  \textbf{\bibinfo{volume}{9}} (\bibinfo{year}{2009}).
\newblock \urlprefix\url{https://doi.org/10.2202/1935-1704.1548}.

\bibitem{Nowak2006}
\bibinfo{author}{Nowak, M.~A.}
\newblock \bibinfo{title}{Five rules for the evolution of cooperation}.
\newblock \emph{\bibinfo{journal}{Science}} \textbf{\bibinfo{volume}{314}},
  \bibinfo{pages}{1560--1563} (\bibinfo{year}{2006}).
\newblock \urlprefix\url{https://doi.org/10.1126/science.1133755}.

\bibitem{Riolo2001}
\bibinfo{author}{Riolo, R.~L.}, \bibinfo{author}{Cohen, M.~D.} \&
  \bibinfo{author}{Axelrod, R.}
\newblock \bibinfo{title}{Evolution of cooperation without reciprocity}.
\newblock \emph{\bibinfo{journal}{Nature}} \textbf{\bibinfo{volume}{414}},
  \bibinfo{pages}{441--443} (\bibinfo{year}{2001}).
\newblock \urlprefix\url{https://doi.org/10.1038/35106555}.

\bibitem{Traulsen2007}
\bibinfo{author}{Traulsen, A.} \& \bibinfo{author}{Nowak, M.~A.}
\newblock \bibinfo{title}{Chromodynamics of cooperation in finite populations}.
\newblock \emph{\bibinfo{journal}{PLoS One}} \textbf{\bibinfo{volume}{2}},
  \bibinfo{pages}{e270} (\bibinfo{year}{2007}).

\bibitem{Antal2009}
\bibinfo{author}{Antal, T.}, \bibinfo{author}{Ohtsuki, H.},
  \bibinfo{author}{Wakeley, J.}, \bibinfo{author}{Taylor, P.~D.} \&
  \bibinfo{author}{Nowak, M.~A.}
\newblock \bibinfo{title}{Evolution of cooperation by phenotypic similarity}.
\newblock \emph{\bibinfo{journal}{Proceedings of the National Academy of
  Sciences}} \textbf{\bibinfo{volume}{106}}, \bibinfo{pages}{8597--8600}
  (\bibinfo{year}{2009}).

\bibitem{Fu2012}
\bibinfo{author}{Fu, F.} \emph{et~al.}
\newblock \bibinfo{title}{Evolution of in-group favoritism}.
\newblock \emph{\bibinfo{journal}{Scientific Reports}}
  \textbf{\bibinfo{volume}{2}} (\bibinfo{year}{2012}).
\newblock \urlprefix\url{https://doi.org/10.1038/srep00460}.

\bibitem{Hadzibeganovic2014}
\bibinfo{author}{Hadzibeganovic, T.}, \bibinfo{author}{Lima, F. W.~S.} \&
  \bibinfo{author}{Stauffer, D.}
\newblock \bibinfo{title}{Benefits of memory for the evolution of tag-based
  cooperation in structured populations}.
\newblock \emph{\bibinfo{journal}{Behavioral Ecology and Sociobiology}}
  \textbf{\bibinfo{volume}{68}}, \bibinfo{pages}{1059--1072}
  (\bibinfo{year}{2014}).
\newblock \urlprefix\url{https://doi.org/10.1007/s00265-014-1718-7}.

\bibitem{Zhang2015}
\bibinfo{author}{Zhang, H.}
\newblock \bibinfo{title}{Moderate tolerance promotes tag-mediated cooperation
  in spatial prisoner's dilemma game}.
\newblock \emph{\bibinfo{journal}{Physica A: Statistical Mechanics and its
  Applications}} \textbf{\bibinfo{volume}{424}}, \bibinfo{pages}{52--61}
  (\bibinfo{year}{2015}).
\newblock \urlprefix\url{https://doi.org/10.1016/j.physa.2015.01.005}.

\bibitem{Garcia2014}
\bibinfo{author}{Garc{\'\i}a, J.}, \bibinfo{author}{van Veelen, M.} \&
  \bibinfo{author}{Traulsen, A.}
\newblock \bibinfo{title}{Evil green beards: Tag recognition can also be used
  to withhold cooperation in structured populations}.
\newblock \emph{\bibinfo{journal}{Journal of theoretical biology}}
  \textbf{\bibinfo{volume}{360}}, \bibinfo{pages}{181--186}
  (\bibinfo{year}{2014}).

\bibitem{Ohtsuki2006}
\bibinfo{author}{Ohtsuki, H.}, \bibinfo{author}{Hauert, C.},
  \bibinfo{author}{Lieberman, E.} \& \bibinfo{author}{Nowak, M.~A.}
\newblock \bibinfo{title}{A simple rule for the evolution of cooperation on
  graphs and social networks}.
\newblock \emph{\bibinfo{journal}{Nature}} \textbf{\bibinfo{volume}{441}},
  \bibinfo{pages}{502--505} (\bibinfo{year}{2006}).

\bibitem{cox1986diffusive}
\bibinfo{author}{Cox, J.~T.} \& \bibinfo{author}{Griffeath, D.}
\newblock \bibinfo{title}{Diffusive clustering in the two dimensional voter
  model}.
\newblock \emph{\bibinfo{journal}{The Annals of Probability}}
  \bibinfo{pages}{347--370} (\bibinfo{year}{1986}).

\bibitem{dornic2001critical}
\bibinfo{author}{Dornic, I.}, \bibinfo{author}{Chat{\'e}, H.},
  \bibinfo{author}{Chave, J.} \& \bibinfo{author}{Hinrichsen, H.}
\newblock \bibinfo{title}{Critical coarsening without surface tension: The
  universality class of the voter model}.
\newblock \emph{\bibinfo{journal}{Physical Review Letters}}
  \textbf{\bibinfo{volume}{87}}, \bibinfo{pages}{045701}
  (\bibinfo{year}{2001}).

\end{thebibliography}

\section*{Author contributions statement}

G.G.J. and S.B. conceived the model. G.G.J. performed the simulations. The authors collaborated in writing the manuscript.

\section*{Additional information}


\paragraph{Accession codes}
All code used to produce the included figures, has been submitted as supplementary material.

\paragraph{Competing interests}
The author(s) declare no competing interests.

\vfill
\pagebreak

\section{Supplementary figures}
All panels in all figures below show parameter scans. 
`label/strat-correlation' measures to what extent agents positively 
discriminate their own label. Add one for every agent who has a 
cooperating strategy towards their own badge, and subtract one 
for each agent cooperating with the label they don't carry themselves.

\subsection{Uniform label distribution}
\begin{center}
	\includegraphics[width=\textwidth]{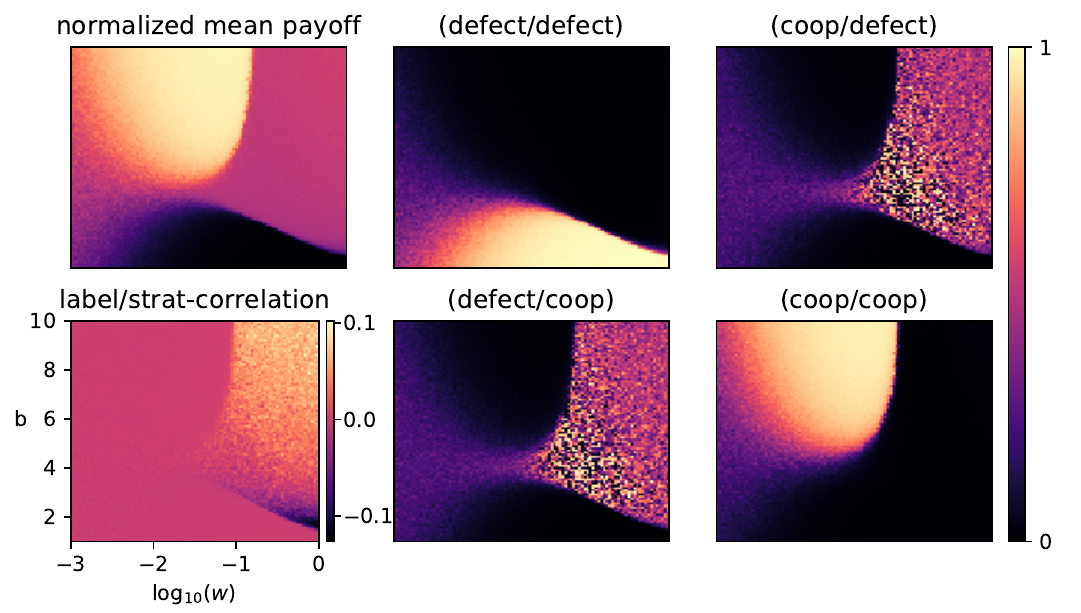}    
\end{center}
The parameter-scan shown here differs from the model described 
in the main text only by having a uniform label distribution.
Every agent has 50/50 percent chance of being blue or green.
The label distribution is random and redrawn from a uniform 
distribution at every data point.
\label{fig:uniform-scan}

\vfill
\pagebreak

\subsection{Random Graphs - average connectivity 4}
\begin{center}
	\includegraphics[width=1.\textwidth]{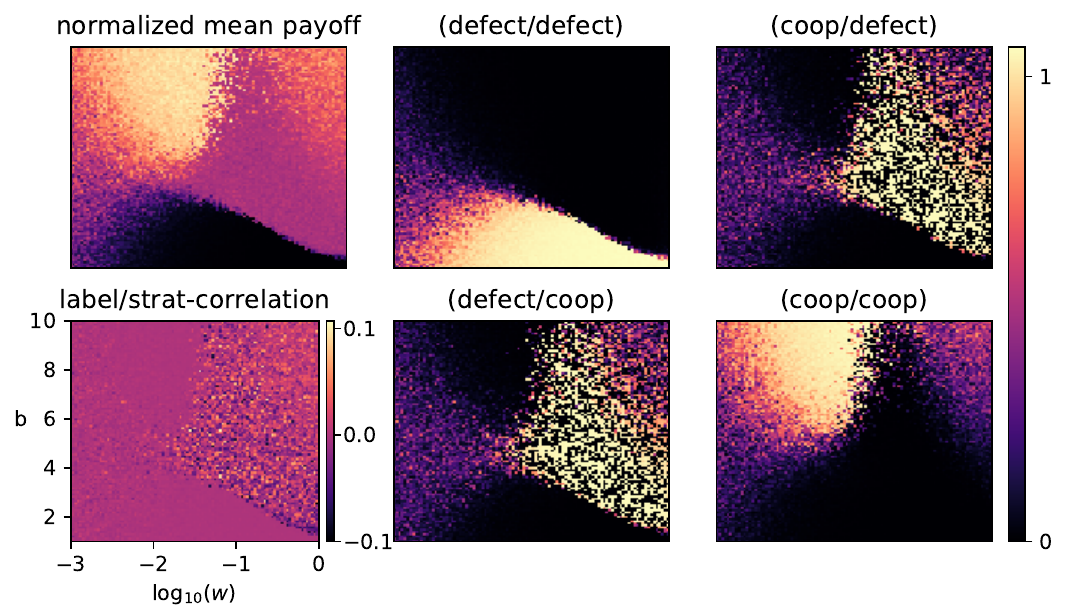}    
\end{center}
As the model described in this paper is defined in terms of local interactions, 
it is straightforward to extent the investigations to arbitrary graph topologies. 
The parameter-scan shown here is executed on binomial random-graphs with $1000$ 
nodes and average connectivity $4$. A new random graph, and a new random uniform
label-distribution is drawn for every time series.
\label{fig:random-graphs}

\vfill
\pagebreak

\subsection{1D - Ring}
\begin{center}
	\includegraphics[width=1.\textwidth]{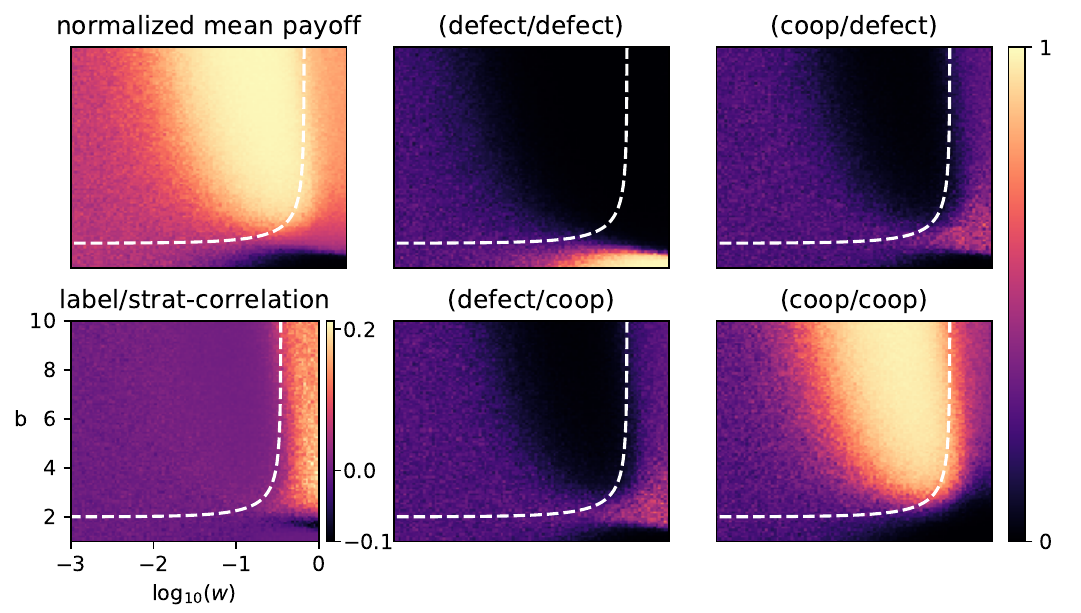}    
\end{center}
For comparison with the theoretically predicted phase-transition mentioned 
in the paper (red line), we here show what the phase diagram looks 
like for a 1D system, i.e.\ a line of agents with two neighbors each, 
closed at the ends to form a ring. The label distribution is random 
and redrawn from a uniform distribution at every data point.
\vspace{11cm}
\label{fig:1d-scan}

\vfill
\pagebreak

\subsection{Timeseries - Converging to stable population fractions}
\begin{center}
	\includegraphics[width=\textwidth]{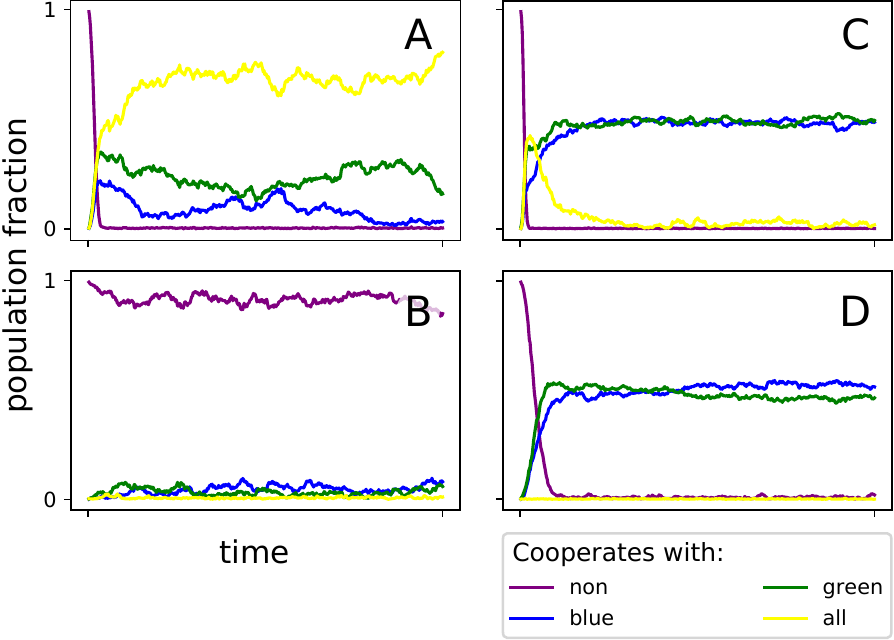}
\end{center}
\label{fig:timeseries}
These four panels show the time-series of the population fraction of each of the four strategies, corresponding to each of the snap-shots in figure \ref{fig:phaseDiagram} in the main text.
Each simulation starts with a population of $100\%$ defectors.
The figures strongly suggest, that the model dynamics reaches a stationary state, with some stochastic fluctuations around stationary strategy-distributions.
\vspace{5cm}

\vfill
\pagebreak

\subsection{Analytic solution of simplified 1D system}

As mentioned in main text, the phase-transition can be analytically calculated in a simplified version of the 1D system with out mutation ($\mu=0$).
\begin{center}
	\includegraphics[width=0.5\linewidth]{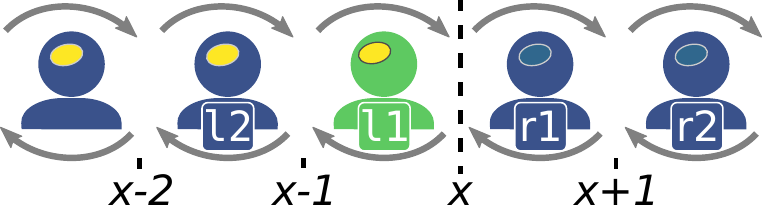}
\end{center}
Imagine a line of agents, each connected to their two nearest neighbours. One agent in the middle of the line has the green label, while all others have the blue label.
The dynamical variable in this system is a point on the line, $x$, such that all agents to the left of that point have the yellow strategy (cooperate with everyone), and all agents to the right have the blue strategy (cooperate only with blue neighbours).
At any point in time, we enumerate the agents according to how far they are from the border. The first agent to the right (left) is called $r1$ ($l1$), the second $r2$ ($l2$), and so on.

Since the mutation is assumed to be zero, the border between the two strategies can only move if the agent immediately at one side copies the strategy of the agent on the other.

The boundary moves left if the first agent left of the boundary ($l1$) imitates the strategy of the first player right of the boundary ($r1$). This happens with probability:
$$ P(x\rightarrow x-1) = \frac{1}{N}\frac{f_{r1}}{f_{r1}+f_{l2}} = \frac{1}{N}\frac{1}{1+f_{l2}/f_{r1}} = \frac{1}{N}\frac{1}{1+\exp{(w\cdot(p_{l2} - p_{r1}))}},$$
given the boundary is at position $x$, where $f_{i}$ and $p_{i}$ are the fitness and payoff of agent $i$, and $l2$ is the second agent left of the boundary.
Likewise the probability that the boundary will move right is:
$$ P(x\rightarrow x+1|x) =  \frac{1}{N}\frac{1}{1+\exp{(w\cdot(p_{r2} - p_{l1}))}},$$%
where $r2$ is the second agent left of the boundary.
After having determined the individual stepping probabilities we can calculate the stationary distribution of the boundary by assuming microscopic balance: 
$$P(x)P(x\rightarrow x+1|x) = P(x)P(x+1\rightarrow x|x+1),
$$
where P(x) is the stationary probability that the boundary is at position $x$.
From this we get an expression for the fraction between the stationary probability of a position $x$  and at a position $x+n$:
$$ \frac{P(x)}{P(x+n)} = 
\frac{P(x+1\rightarrow x|x+1)}{P(x\rightarrow x+1|x)}\cdot
\frac{P(x+2\rightarrow x+1|x+2)}{P(x+1\rightarrow x+2|x+1)}\cdot
...\cdot
\frac{P(x+n\rightarrow x+n-1|x+n)}{P(x+n-1\rightarrow x+n|x+n-1)}.
$$
If we assume $x$ is a position at least three steps to the completely cooperative side of the green agent (the discriminating strategy dominates) and $x+n$ is a position at least three steps to the discriminating side (the completely cooperative strategy dominates) then we obtain the following expression after reducing factors which appear in both numerator and denominator.
$$ \frac{\mathit{P}(\text{discrimination dominates})}{\mathit{P}(\text{full cooperation dominates})} = 
\frac{1+e^{-2wb}}{2}\frac{(1+e^{w})^2}{(1+e^{-w})^2} =
\frac{e^{2w}}{\tanh{wb}+1}
$$
When the fraction is greater than one, we expect a strategy of asymmetric cooperation (i.e. discrimination) to be able to outperform the full cooperation in a mixed population. Comparison with the 
phase diagram for the one-dimensional system show a close, but non-perfect match 
(see supplementary figure~\ref{fig:1d-scan}). 
In the limit of very high cooperation benefit, $b\rightarrow\infty$, this condition reduces to:
$$  w > \log{2}/2
$$
This result shows, that no matter how high the cooperation-benefit is, there will always be a critical selection-pressure above which discriminating strategies are more likely to spread than unconditional cooperation.

\end{document}